# Impact of decoherence on the fidelity of quantum gates leaving the computational subspace

Tahereh Abad[1], Yoni Schattner[2], Anton Frisk Kockum[1], and Göran Johansson[1]

[1]Department of Microtechnology and Nanoscience, Chalmers University of Technology, 412 96 Gothenburg, Sweden
[2]AWS Center for Quantum Computing, Pasadena, CA 91125, USA

**The fidelity of quantum operations is often limited by incoherent errors, which typically can be modeled by fundamental Markovian noise processes such as amplitude damping and dephasing. In Phys. Rev. Lett. 129, 150504 (2022), we presented an analytical result for the average gate fidelity of a general multiqubit operation in terms of the dissipative rates and the corresponding Lindblad jump operators, provided that the operation remains in the computational subspace throughout the time evolution. Here we generalize this expression for the average gate fidelity to include the cases where the system state temporarily leaves the computational subspace during the gate. Such gate mechanisms are integral to several quantum-computing platforms, and our formula is applicable to all of them; as examples, we employ it for the two-qubit controlled-Z gate in both superconducting qubits and neutral atoms. We also obtain the average gate fidelity for simultaneous operations applied in multiqubit systems. These results are useful for understanding the error budgets of quantum gates while scaling up quantum computers.**

## 1 Introduction

Architectures such as circuit quantum electrodynamics [1–4], trapped ions [5, 6], quantum dots [7], and photonics [8] present promising paths to building a quantum computer that has the potential to solve problems that are classically intractable [1, 9–17]. To reach this goal, the ability to implement high-fidelity quantum operations is essential. For example, achieving quantum control with high fidelity lies at the heart of enabling fault-tolerant quantum computing [12, 18–21].

Tahereh Abad: Tahereh.Abad@chalmers.se

For fault-tolerant computation, characterizing and reducing the remaining errors becomes increasingly challenging [22, 23]. Quantum process tomography can completely characterize a gate, decomposing a process into Pauli or Kraus operators [24]. However, improving gates is complicated: gate parameters map non-intuitively onto the process matrix, and state preparation and measurement errors can be confused with process errors. The well-understood approach to achieve high-fidelity gates, Clifford-based randomized benchmarking (RB) [25–28], maps gate errors onto control parameters and feeds this back to optimize the gates.

In Ref. [29], a metrological tool based on RB to quantify noise on time scales relevant for quantum gates was introduced. That work included an analytical expression for the effect of noise during an idle gate period in an RB sequence. However, this expression, which has been used in several experimental studies since [30, 31], was only derived for single-qubit Clifford gates. Since single-qubit gates now can be performed with very high fidelity, the focus of recent experimental work is on improving two-qubit gates [32–36], a much more challenging task. Furthermore, by controlling multiple two-qubit couplings simultaneously [37, 38], three-qubit iToffoli gates [39] and fast three-qubit controlled-CPHASE-SWAP (CCZS) gates [40] have been implemented.

To improve the performance of quantum operations, understanding the effect of decoherence such as amplitude damping (energy relaxation) and dephasing on both single- and multi-qubit gates is essential. In Refs. [41, 42] analytical results for quantifying the effect of decoherence on fidelity have been given, provided that the quan-

Accepted in ⟨Quantum 2025-03-01, click title to verify. Published under CC-BY 4.0. 1

tum operations take place in the computational subspace, i.e., the states $\{|0\rangle, |1\rangle\}$ of the qubits; dissipation leading to leakage to states outside of the computational subspace, e.g., heating processes, can still be accounted for.

Since many quantum gates, in various quantum-computing architectures, rely on mechanisms that temporarily populate states outside the computational subspace, a natural extension of the work in Refs. [41, 42] is to account also for such processes. A typical example of a gate going outside the computational subspace is the two-qubit controlled-Z (CZ) gate, where in superconducting circuits [36, 43, 44] a full swap between $|11\rangle$ and $|02\rangle$ (or $|20\rangle$) and back is used to imprint a $\pi$ phase shift on $|11\rangle$; simultaneous such CZ gates yield the three-qubit CCZS gate [37, 40]. A similar approach is used for realizing multi-qubit entangling gates between individual neutral atoms [45, 46] and trapped ions [47] through Rydberg interactions, where a transition to the Rydberg level, a state outside of the computational subspace, is used.

In this Letter, we derive analytical results for how quantum operations are affected by decoherence, without being limited by whether the time evolution includes transitions to states outside of the computational subspace. We consider the case where errors are dominated by the common processes of energy relaxation and dephasing, acting independently on the individual qubits. Using a Lindblad-master-equation method, we find a simple formula that is applicable to any quantum system. For the CZ and CCZS gates in superconducting qubits, the formula gives a very minor modification of previous results [40, 44, 48], which assumed trace preservation in the computational subspace. We also employ our formula to CZ gates in neutral atoms, simultaneous gates in a multiqubit system, and show how to find a total gate fidelity from the fidelity of individual gates. Our results provide bounds that allow for robust estimation and optimization of gate fidelities across quantum-computing platforms.

Having the easy access to estimates of average gate fidelities from decoherence rates that our results provide opens up for many applications. For example, if the average gate fidelity can be estimated, that can in turn be used to place an upper bound on errors and assess progress towards fault-tolerant quantum computation [49]. Similarly, an estimate of average gate fidelity can in turn be used to estimate whether the quantum approximate optimization algorithm (QAOA) [16] can be used for solving various instances of combinatorial optimization problems [50]. Furthermore, since our results enable calculating the impact of incoherent errors on average gate fidelity, it can be combined with measurements of such gate fidelity to make an error budget that includes the level of coherent errors. Reference [51] then shows how to analyze the robustness of quantum algorithms against coherent errors and presents worst-case fidelity bounds for quantum circuits affected by such errors. Also, average gate fidelities calculated from decoherence rates enables checking whether a quantum computer can be simulated classically [52] and choosing the most suitable gate set for running QAOA [53].

## 2 Average gate fidelity

The average gate fidelity $\overline{F}$, defined as [18]

$$\overline{F} \equiv \int d|\psi\rangle \left\langle \Psi \middle| \hat{U}_g^\dagger \mathcal{E}(|\Psi\rangle\langle\Psi|) \hat{U}_g \middle| \Psi \right\rangle, \quad (1)$$

measures the overlap between the state evolved by the quantum channel $\mathcal{E}$ and the ideal unitary gate operation $\hat{U}_g$. Here, the integral is over all pure initial states $|\psi\rangle$ in the computational subspace and the initial state $|\Psi\rangle = |\psi\rangle \oplus |\mathbf{O}\rangle$, where $|\mathbf{O}\rangle$ is a zero vector in the space of states outside the computational subspace, has no component outside the computational subspace. Note that the evolution operation $\mathcal{E}$ can take states outside the computational subspace, and that $\mathcal{E}$ thus may not preserve the trace in the computational subspace. For $\mathcal{E}$ perfectly implementing $\hat{U}_g$, we have $\overline{F} = 1$.

The gate operation in Eq. (1) can be generated by a time-dependent Hamiltonian $\hat{H}(t)$ applied for a time $\tau$, such that $\hat{U}_g = \hat{U}(\tau) = \mathcal{T} \exp[-\frac{i}{\hbar}\int_0^\tau \hat{H}(t)dt]$, where $\mathcal{T}$ is the time-ordering operator. Adding $N_L$ different dissipative processes, the time evolution of the system is then given by the master equation

$$\dot{\hat{\rho}}(t) = -\frac{i}{\hbar}\left[\hat{H}(t), \hat{\rho}(t)\right] + \sum_{k=1}^{N_L} \Gamma_k \mathcal{D}[\hat{L}_k]\hat{\rho}(t), \quad (2)$$

where $\mathcal{D}[\hat{L}]\hat{\rho} = \hat{L}\hat{\rho}\hat{L}^\dagger - \frac{1}{2}\{\hat{L}^\dagger\hat{L}, \hat{\rho}\}$ is the standard Lindblad superoperator [54], and each process has a corresponding rate $\Gamma_k$ and Lindblad jump operator $\hat{L}_k$. Note that both the ideal gate evolution and the jump operators are allowed to take the system out of the computational subspace.

To describe the weakly dissipative dynamics of the system [35–37, 39, 43, 55–60], one can expand the solution to the master equation in the small parameter $\Gamma_k\tau \ll 1$ [61], and find that each dissipative process contributes independently to $\bar{F}$ to first order in $\Gamma_k\tau$ [41]:

$$\bar{F} = 1 + \sum_{k=1}^{N_L} \Gamma_k \int_0^\tau dt\, \delta F(t, \hat{L}_k) + \mathcal{O}\Big(\tau^2\Gamma_k^2\Big), \quad (3)$$

where

$$\delta F(t, \hat{L}) = \int d\psi \,\mathrm{Tr}\left[\hat{L}^\dagger \hat{\rho}_\psi(t) \hat{L} \hat{\rho}_\psi(t)\right] - \int d\psi \,\mathrm{Tr}\left[\hat{L}^\dagger \hat{L} \hat{\rho}_\psi(t)\right]. \quad (4)$$

Here $\hat{\rho}_\psi(t) = \hat{U}(t)|\psi\rangle\langle\psi|\hat{U}^\dagger(t)$ is the result of the unitary transformation that preserves the purity of the state.

We rewrite Eq. (4) as

$$\int d\psi \Big(\mathrm{Tr}\left[\hat{L}^\dagger(t)\rho_\psi \hat{L}(t)\rho_\psi\right] - \mathrm{Tr}\left[\hat{L}^\dagger(t)\hat{L}(t)\rho_\psi\right]\Big) = \delta F(t, \hat{L}) \equiv \delta F(\hat{L}(t)), \quad (5)$$

where $\hat{L}(t) = \hat{U}^\dagger(t)\,\hat{L}\,\hat{U}(t)$, to be able to use a certain expansion of the density matrix in our calculations later. Here, we note that the trace operation is over the full Hilbert space. Still, since the initial-state density matrix $\rho_\psi$ only has elements in the computational subspace, it also implies a projection of the operators $\hat{L}(t)$, $\hat{L}^\dagger(t)$, and $\hat{L}^\dagger(t)\hat{L}(t)$ onto the computational subspace. The gate fidelity will thus depend explicitly on both the jump operator $\hat{L}$ and the gate $\hat{U}(t)$.

To perform the integral over the initial states for the general $N$-qubit case in Eq. (5), we expand the density matrix as $\hat{\rho} = \frac{1}{d}(\hat{1}_N + \sum_{i=1}^{d^2-1} c_i \hat{f}_i)$, where the $\hat{f}_i$ are tensor products of Pauli matrices (the $N$ indices $i_1 \dots i_N$ are collected into the single combined index $1 \le i \le d^2 - 1$, where $d = 2^N$). Following the symmetry arguments given in Ref. [62], we use $\langle c_i \rangle = 0$ and $\langle c_i c_j \rangle = \delta_{ij}/(d+1)$ [41] and average over all initial states $|\psi\rangle$ to find that the fidelity reduction for the $N$-qubit case becomes (see Appendix A)

$$\delta F_N(\hat{L}(t)) = \frac{1}{d(d+1)} \mathrm{Tr_{cmp}}\Big[\hat{L}^\dagger(t)\hat{\mathbb{1}}_{\mathrm{cmp}}\hat{L}(t)\hat{\mathbb{1}}_{\mathrm{cmp}}\Big] + \frac{1}{d^2(d+1)} \sum_{i=0}^{d^2-1} \mathrm{Tr_{cmp}}\Big[\hat{L}^\dagger(t)\hat{f}_i\hat{L}(t)\hat{f}_i\Big] - \frac{1}{d}\mathrm{Tr_{cmp}}\Big[\hat{L}^\dagger(t)\hat{L}(t)\Big], \quad (6)$$

where "cmp" denotes that the trace is over the states in the computational subspace. Here $\hat{\mathbb{1}}_{\mathrm{cmp}} = \hat{1}_N \oplus \hat{0}$ is the identity operation applied to the computational subspace of N qubits, but without support outside the computational subspace ($\hat{0}$ is a zero matrix in the space outside the computational subspace), and therefore $\mathrm{Tr_{cmp}}[\hat{L}^\dagger(t)\mathbb{1}_{\mathrm{cmp}}\hat{L}(t)] \neq \mathrm{Tr_{cmp}}[\hat{L}^\dagger(t)\hat{L}(t)]$, i.e., the first and last terms of the right-hand side of Eq. (6) cannot be added directly. Note that since the unitary operation might take the state outside of the computational subspace, $\mathrm{Tr_{cmp}}\,[\hat{L}^\dagger(t)\hat{L}(t)] \neq \mathrm{Tr_{cmp}}\,[\hat{L}^\dagger\hat{L}]$.

To simplify Eq. (6) further, we project the time-dependent jump operator $\hat{L}(t)$ into the computational subspace. Representing it in the $\hat{f}_i$ basis, we obtain terms like $\sum_i \mathrm{Tr}\,[\hat{f}_j\, \hat{f}_i\, \hat{f}_k\, \hat{f}_i]$, where indices $j$ and $k$ are associated with contributions from $\hat{L}^\dagger(t)$ and $\hat{L}(t)$, respectively. In Appendix B, we show that $\sum_{i=0}^{d^2-1} \mathrm{Tr}\,[\hat{f}_j\, \hat{f}_i\, \hat{f}_k\, \hat{f}_i] = d^3\delta_{j0}\delta_{k0}$, so the only nonzero term of the summation in Eq. (6) is the contribution of the identity $\hat{f}_0$ in $\hat{L}(t)$. Thus, the fidelity reduction for the $N$-qubit operation reduces to (see Appendix B)

$$\delta F_N(\hat{L}(t)) = \frac{1}{d(d+1)} \mathrm{Tr_{cmp}}\left[\hat{L}^\dagger(t)\right] \mathrm{Tr_{cmp}}\left[\hat{L}(t)\right] + \frac{1}{d(d+1)} \mathrm{Tr_{cmp}}\Big[\hat{L}^\dagger(t)\hat{\mathbb{1}}_{\mathrm{cmp}}\hat{L}(t)\hat{\mathbb{1}}_{\mathrm{cmp}}\Big] - \frac{1}{d}\mathrm{Tr_{cmp}}\left[\hat{L}^\dagger(t)\,\hat{L}(t)\right]. \quad (7)$$

This is *the main result* of this article. Together with Eq. (3), it means that the average gate fidelity depends on the operation time, the dissipation rate, and the time-evolved jump operator. Note that $d = 2^N$ no matter how many levels each qubit has beyond its computational subspace.

In addition to the fidelity, it is often useful to characterize processes by the leakage outside of the computational subspace, defined as $\bar{\mathcal{L}} \equiv \int d|\psi\rangle\,(1 - \mathrm{Tr_{cmp}}\,[\mathcal{E}(|\psi\rangle\langle\psi|)])$ [63]. Following similar steps as those leading to Eqs. (3)

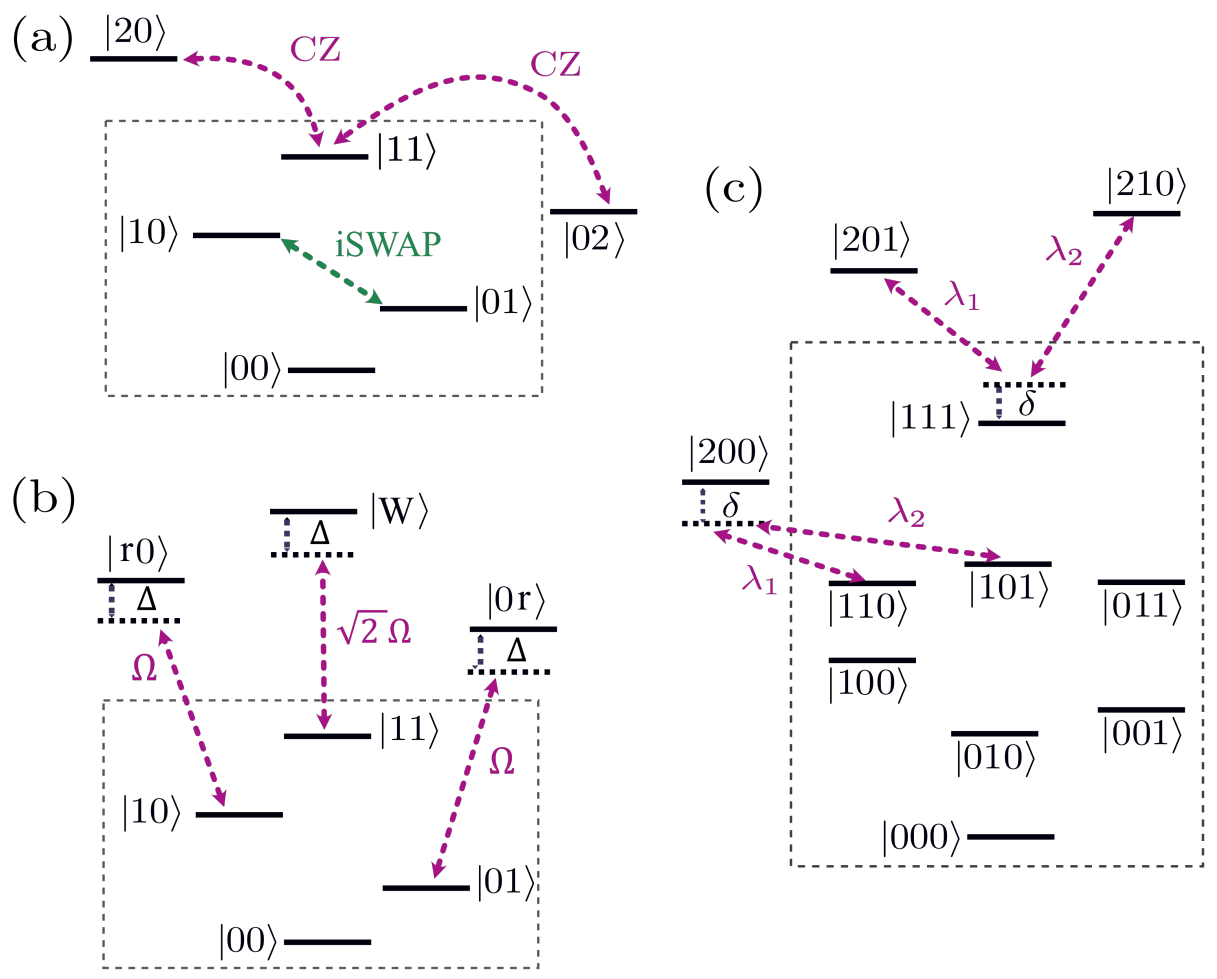

Figure 1: Examples of gate operations using states outside the computational subspace. (a) In superconducting qubits, iSWAP gates are confined to the computational subspace (dashed rectangle), but some CZ gates are not. (b) Transitions for a CZ gate with neutral atoms. (c) Transitions for a CCZS gate with superconducting qubits.

and (7), we obtain $\bar{\mathcal{L}} = \sum_{k=1}^{N_L} \Gamma_k \int_0^\tau dt\, \delta\mathcal{L}(t, \hat{L}_k) + \mathcal{O}(\tau^2\Gamma_k^2)$, with

$$\delta\mathcal{L}(t, \hat{L}) = \frac{1}{d}\mathrm{Tr}_{\mathrm{cmp}}\left[\hat{L}^\dagger(t)\hat{L}(t)\right] - \frac{1}{d}\mathrm{Tr}_{\mathrm{cmp}}\left[\hat{L}^\dagger(t)\hat{\mathbb{1}}_{\mathrm{cmp}}\hat{L}(t)\right]. \tag{8}$$

## 3 Operations in the computational subspace

For quantum operations confined to the computational subspace, e.g., a two-qubit iSWAP gate ($|01\rangle/|10\rangle \to i|10\rangle/|01\rangle$), as shown in Fig. 1(a), or a three-qubit iToffoli gate ($|110\rangle \to i|111\rangle$), $\hat{U}(t)$ only performs a rotation in the Hilbert space that the trace is taken over, so using $\hat{U}(t)\hat{U}(t)^\dagger = 1$, Eq. (7) becomes

$$\delta F_{\mathrm{cmp}}(\hat{L}) = \frac{1}{d(d+1)}\,\mathrm{Tr}\left[\hat{L}^\dagger\right]\mathrm{Tr}\left[\hat{L}\right] - \frac{1}{d+1}\,\mathrm{Tr}\left[\hat{L}^\dagger\,\hat{L}\right]. \tag{9}$$

The same result can be obtained considering a qudit ($d$-level system) instead of an $N$-qubit system [64]. This fidelity reduction is independent of the specific operation; it depends only on the operation time and the dissipation. Note that just as the ideal gate evolution, the jump operators are here confined to the computational subspace. For energy relaxation acting on one qubit with jump operator $\hat{\sigma}_-$ and rate $\Gamma_1$, or pure dephasing with jump operator $\hat{\sigma}_z$ and rate $\Gamma_\phi$ [the rate multiplying the dissipator in Eq. (2) is $\Gamma_\phi/2$, making the coherence decay with the rate $\Gamma_\phi$], we obtain

$$\delta F_{\mathrm{cmp}}(\hat{\sigma}_z \otimes \hat{1}_{N-1}) = 2\,\delta F_{\mathrm{cmp}}(\hat{\sigma}_- \otimes \hat{1}_{N-1}) = -\frac{d}{d+1}, \tag{10}$$

in agreement with the results in Ref. [41].

We note that, following the argument given in Ref. [41], when we apply a quantum operation, irrespective of remaining in or leaving the computational subspace, coherent errors and incoherent errors contribute independently to the fidelity reduction to the first order. We now proceed to evaluate the fidelity reduction in the presence of energy relaxation and dephasing for a few relevant quantum operations leaving the computational subspace.

## 4 CZ gates with superconducting qubits

Two-qubit gates available in the transmonlike [65] superconducting qubit architecture include (variations on) CPHASE (controlled-phase), iSWAP, and $\sqrt{\mathrm{iSWAP}}$ gates [66]. While the latter ones are confined to the computational subspace and are covered by Eq. (10), a CZ gate created by swapping between $|11\rangle$ and $|20\rangle$ (through the Hamiltonian $\hat{H}_{\mathrm{CZ}} = \lambda\left(|11\rangle\langle 20| + |20\rangle\langle 11|\right)$) is not; see Fig. 1(a). The unitary operation $\hat{U}_{\mathrm{CZ}}(t) = \exp[-i\hat{H}_{\mathrm{CZ}}t]$ adds a phase factor of $-1$ to $|11\rangle$ in the gate time $\tau = \pi/\lambda$.

For a three-level transmon, energy relaxation is described by jump operator $\hat{L}_- = \hat{\sigma}_{01}^- + \sqrt{2}\hat{\sigma}_{12}^-$ and rate $\Gamma_1$. For the two three-level transmons involved in a CZ gate, we descibe the effect of energy relaxation acting on both qubits individually by $\hat{L}_-^{q1} = \hat{L}_- \otimes \hat{1}$ and $\hat{L}_-^{q2} = \hat{1} \otimes \hat{L}_-$, with the rates $\Gamma_1^{q1}$ and $\Gamma_1^{q2}$, respectively. These jump operators are traceless, and their evolution by $\hat{U}(t)^\dagger$ is still traceless because that unitary is a change of basis, so the first term in Eq. (7) vanishes. The time-evolved jump operator corresponding to energy relaxation on qubit 1 becomes (see Appendix C)

$$\begin{aligned}\hat{L}_{-}^{q1}(t) &= |00\rangle\langle 10| \\ &+ \cos(\lambda t)\Big(|01\rangle\langle 11| + \sqrt{2}|10\rangle\langle 20|\Big) \\ &- i\sin(\lambda t)\Big(|01\rangle\langle 20| + \sqrt{2}|10\rangle\langle 11|\Big). \quad (11)\end{aligned}$$

It is straightforward to calculate the remaining terms in Eq. (7) and obtain $\int_0^\tau dt\, \delta F_2(\hat{L}_{-}^{q1}(t)) = -\frac{1}{2}\tau$ (see Appendix C).

Transmon dephasing is described by the jump operator $\sum_{j=1}^{d} 2j\, |j\rangle\langle j|$ and rate $\Gamma_\phi/2$, or $\hat{L}_\phi = |1\rangle\langle 1| + 2\,|2\rangle\langle 2|$ with rate $2\,\Gamma_\phi$. This process acts on both qubits individually through $\hat{L}_\phi^{q1} = \hat{L}_\phi \otimes \hat{1}$ and $\hat{L}_\phi^{q2} = \hat{1} \otimes \hat{L}_\phi$, with the rates $2\Gamma_\phi^{q1}$ and $2\Gamma_\phi^{q2}$, respectively. In a similar fashion as for energy relaxation, we find the fidelity reduction due to dephasing and then add up all these contributions to obtain the average gate fidelity for the CZ gate (see Appendix C)

$$\overline{F}_{\mathrm{CZ}} = 1 - \frac{1}{2}\Gamma_1^{q1}\,\tau - \frac{3}{10}\Gamma_1^{q2}\,\tau - \frac{31}{40}\Gamma_\phi^{q1}\,\tau - \frac{3}{8}\Gamma_\phi^{q2}\,\tau. \quad (12)$$

This result is slightly different from Refs. [44, 48] (by a fraction 1/80 in the terms with pure dephasing), where they used Ref. [18] to calculate the average gate fidelity. That calculation requires the evolution to be trace-preserving in the computational subspace, which is not the case for some part of the pure dephasing processes at work in this example. We also go beyond those and also find a general formula for an imperfect CZ gate, where $\lambda \neq \pi/\tau$ (coherent error); see Appendix C.

We note in Eq. (12) that since qubit 1 populates $|2\rangle$, it is more strongly affected by relaxation ($-\Gamma_1^{q1}\,\tau/2$) than when it is confined to the computational subspace [$-2\,\Gamma_1^{q1}\,\tau/5$ in Eq. (10)], whereas energy relaxation on qubit 2 matters less ($-3\,\Gamma_1^{q1}\,\tau/10$) since the gate operation tends to keep that qubit in $|0\rangle$. If there is a choice between which of two qubits in a two-qubit gate that should leave the computational subspace, it is thus best to choose the one more robust to decoherence.

## 5 CZ gates with neutral atoms

Numerous protocols for entangling atoms using Rydberg interactions have been explored theoretically and experimentally [67–71]. A relevant entangling gate between atoms is the CZ gate that maps the computational basis states as $|00\rangle \to |00\rangle$, $|01\rangle \to |01\rangle e^{i\phi}$, $|10\rangle \to |10\rangle e^{i\phi}$ and $|11\rangle \to |11\rangle e^{i(2\phi+\pi)}$, up to a single-qubit phase. Qubits are encoded in long-lived hyperfine states $\{|0\rangle, |1\rangle\}$; realization of the CZ map relies on the Rydberg blockade, implemented by coupling $|1\rangle$ to the Rydberg state $|r\rangle$ [45, 46]. This is done by applying two laser pulses, each of length $\tau$ at detuning $\Delta$ and Rabi frequency $\Omega$, with a phase jump $\xi$ in between.

The state $|00\rangle$ is uncoupled to the dynamics and does not change. For $|01\rangle$ and $|10\rangle$, where one of the atoms is initially in state $|0\rangle$ and remains unchanged, the other atom evolves through

$$\hat{H}_1 = \frac{1}{2}(\Omega|1\rangle\langle r| + \Omega^*|r\rangle\langle 1|) - \Delta|r\rangle\langle r|. \quad (13)$$

If both atoms are initially in state $|1\rangle$, under the Rydberg-blockade constraint, where Rydberg-Rydberg interaction is much larger than $|\Omega|$ and $|\Delta|$, the state $|11\rangle$ is coupled to $|W\rangle = \frac{1}{\sqrt{2}}(|r1\rangle + |1r\rangle)$ through the Hamiltonian

$$\hat{H}_2 = \frac{\sqrt{2}}{2}(\Omega|11\rangle\langle W| + \Omega^*|W\rangle\langle 11|) - \Delta|W\rangle\langle W|. \quad (14)$$

The transitions described here are shown in Fig. 1(b).

To apply the CZ gate, $\Delta$ is fixed and $\tau$ is selected such that the first pulse drives a perfect detuned Rabi oscillation on $|11\rangle$ (with enhanced Rabi frequency $\sqrt{2}\Omega$) and then an incomplete detuned Rabi oscillation for $|01\rangle$ (with Rabi frequency $\Omega$). The phase $\xi$ of the second pulse corresponds to driving the system around a different axis on the Bloch sphere constructed by $|01\rangle$ and $|0r\rangle$ (or $|11\rangle$ and $|W\rangle$) as north and south pole. By tuning $\xi$ it is possible to close the trajectory of $|01\rangle$ while driving a second complete cycle for $|11\rangle$, such that both $|01\rangle$ and $|11\rangle$ return to their initial positions on the Bloch sphere with accumulated dynamical phases $\phi_{01}$ and $\phi_{11}$, respectively. Choosing $\Delta$ to obtain $\phi_{11} = 2\phi_{01} - \pi$ yields the CZ gate.

A Rydberg state is always subject to energy relaxation, but only a small fraction is relaxation to the computational subspace; the remaining decay is transitions to nearby Rydberg states or the ground state [72]. In gates using Rydberg interactions, knowing the effect of dissipation on the Rydberg state during gates is thus essential. We consider the jump operator $\hat{L}_r = |\mathbf{O}\rangle\langle r|$, describing energy relaxation to states $|\mathbf{O}\rangle$ outside the computational subspace, and individual decay rates $\Gamma_r^{q1}$ and $\Gamma_r^{q2}$. Using Eq. (7) and that the system is symmetric under permutation of the atoms and their decay rates, we obtain (see Appendix D)

$$\overline{F}_{\mathrm{RydbergCZ}} = 1 - \frac{15}{58}\left(\Gamma_r^{q1} + \Gamma_r^{q2}\right)\tau. \tag{15}$$

For example, in a $^{171}$Yb neutral atom, for $n = 75$, the decay rate to nearby Rydberg states and the ground state are 3480 $s^{-1}$ and 1918 $s^{-1}$, respectively, which gives a total rate of $\Gamma_r = 5398\ s^{-1}$ [72]. Considering the total time $2\tau \approx 2.732\,\pi/\Omega$ with effective Rabi frequency $\Omega \approx 2\pi \times 3.5$ MHz, we have $\Gamma_r\tau = 0.001$, leading to an upper bound $\overline{F}_{\mathrm{RydbergCZ}} = 0.9994$, whereas the measured gate fidelity is $\geq 0.974(3)$ [45].

## 6 Three-qubit gates

A three-qubit CCZS gate consists of two simultaneous CZ gate operations on qubit pairs $(q_1, q_2)$ and $(q_1, q_3)$, with strengths $\lambda_1$ and $\lambda_2$, both detuned by $\Delta$, with $q_1$ the qubit where the second excited state $|2\rangle$ is populated during the gate [37, 40]; the relevant transitions are shown in Fig. 1(c). With the three-qubit states denoted $|q_1q_2q_3\rangle$, the Hamiltonian is

$$\begin{aligned}\hat{H} = &[\lambda_1(t)(|110\rangle\langle 200| + |111\rangle\langle 201|) \\ &+ \lambda_2(t)(|101\rangle\langle 200| + |111\rangle\langle 210|) + \text{H.c.}] \\ &+ \delta(|200\rangle\langle 200| - |111\rangle\langle 111|),\end{aligned} \tag{16}$$

where H.c. denotes Hermitian conjugate. This Hamiltonian results in applying both CZ and SWAP gates to $(q_2, q_3)$ conditioned on $q_1$ being in $|1\rangle$ [37, 40] (see Appendix E). We evaluate the average gate fidelity for a subclass of CCZS gates: $\lambda_1 = \lambda, \lambda_2 = -\lambda e^{i\varphi}$, $\delta = 0$, and gate time $\tau = \pi/\sqrt{2}\lambda$, for which we find (see Appendix E)

$$\begin{aligned}\overline{F}_{\mathrm{CCZS}} = 1 &- \frac{163}{288}\Gamma_1^{q1}\tau - \frac{7}{18}\left(\Gamma_1^{q2} + \Gamma_1^{q3}\right)\tau \\ &- \frac{41}{48}\Gamma_\phi^{q1}\tau - \frac{85}{192}\left(\Gamma_\phi^{q2} + \Gamma_\phi^{q3}\right)\tau,\end{aligned} \tag{17}$$

which is $\varphi$-independent. Assuming $\lambda_1 = \lambda_2 = \lambda$ (i.e., $\varphi = \pi$) and $\delta = 0$, Ref. [40] has slightly different pre-factors for the first relaxation term and the pure dephasing terms in Eq. (17), since they used Ref. [18], which assumes that the trace is preserved in the computational subspace.

## 7 Simultaneous gates

In an $N$-qubit system with independent decoherence processes acting on the individual qubits, many different combinations of multi-qubit gates can be applied in parallel. Here we show how to calculate the total average gate fidelity of the whole system by considering one of the parallel gates at a time. As each decoherence channel contributes independently, proportionally to its rate $\Gamma$ and the factor $\delta F(\hat{L}(t))$ [see Eq. (3)], the effect of decoherence on the qubits that are involved in one $m$-qubit gate, while the other qubits are evolving by their own perfect gates, is given by $\hat{L} = \hat{L}_m \otimes \hat{1}_{N-m}$. Therefore $\mathrm{Tr}_{\mathrm{cmp}}[\hat{L}(t)] = \mathrm{Tr}_{\mathrm{cmp}}[\hat{L}_m(t)] \times \mathrm{Tr}[\hat{1}_{N-m}]$ and Eq. (7) for the simultaneous gates reduces to

$$\begin{aligned}\delta F_{m,N}(\hat{L}(t)) &= \delta F_N(\hat{L}_m(t) \otimes \hat{1}_{N-m}) \\ &= \frac{d}{d_m^2(d+1)}\mathrm{Tr}_{\mathrm{cmp}}\left[\hat{L}_m^\dagger(t)\right]\mathrm{Tr}_{\mathrm{cmp}}\left[\hat{L}_m(t)\right] \\ &+ \frac{1}{d_m(d+1)}\mathrm{Tr}_{\mathrm{cmp}}\left[\hat{L}^\dagger(t)\hat{\mathbb{1}}_{\mathrm{cmp}}\hat{L}(t)\hat{\mathbb{1}}_{\mathrm{cmp}}\right] \\ &- \frac{1}{d_m}\mathrm{Tr}_{\mathrm{cmp}}\left[\hat{L}_m^\dagger(t)\,\hat{L}_m(t)\right],\end{aligned} \tag{18}$$

where $d_m = 2^m$ is the dimension of the subsystem we focus on. When $m = N$, Eq. (18) reduces to Eq. (7). As an example, we consider a four-qubit system where simultaneous CZ gates are applied to the qubit pairs $(q_1, q_2)$ and $(q_3, q_4)$. Reusing calculations leading to Eq. (12) for the CZ gate, we find (see Appendix F)

$$\begin{aligned}\overline{F}_{\text{CZ-CZ}} = 1 &- \frac{10}{17}\left(\Gamma_1^{q1} + \Gamma_1^{q3}\right)\tau \\ &- \frac{6}{17}\left(\Gamma_1^{q2} + \Gamma_1^{q4}\right)\tau \\ &- \frac{245}{272}\left(\Gamma_\phi^{q1} + \Gamma_\phi^{q3}\right)\tau \\ &- \frac{117}{272}\left(\Gamma_\phi^{q2} + \Gamma_\phi^{q4}\right)\tau,\end{aligned} \tag{19}$$

where we note that the fidelity reduction is larger (smaller) for qubit 1 (2) compared to the result for the CCZS gate in Eq. (17).

## 8 Conclusion and outlook

We investigated the effect of weak decoherence on the fidelity of arbitrary quantum operations, including the cases where states outside the computational subspace are populated during the time evolution. We derived a simple formula, in terms of dissipative rates and the corresponding Lindblad jump operators, covering all these cases. The formula can be applied to any quantum-computing platform as a powerful tool to help estimate and optimize gate fidelities, and, by extension, the computational power of the noisy quantum hardware. We illustrated this applicability by using the formula to calculate average gate fidelities for two-qubit CZ gates on different platforms and the three-qubit CCZS gate. We also showed how to combine our results to compute the total average gate fidelity for several multi-qubit gates executed in parallel in a larger system by considering each multi-qubit gate separately.

Since the average gate fidelity for operations that venture outside the computational subspace depends on the time evolution for the particular operation in question, some follow-up work is needed to apply our general formula to specific quantum operations in various quantum-computing architectures beyond the examples we have presented here. An interesting case would be operations for continuous-variable quantum computation with superconducting microwave circuits, where logical qubit states are encoded in bosonic systems.

We thank Jorge Fernández-Pendás for useful discussions. This work is unrelated to YS's work at Amazon. We acknowledge support from the Knut and Alice Wallenberg Foundation through the Wallenberg Centre for Quantum Technology (WACQT), from the EU Flagship on Quantum Technology H2020-FETFLAG-2018-03 project 820363 OpenSuperQ and the Horizon Europe programme HORIZON-CL4-2022-QUANTUM-01-SGA via the project 101113946 OpenSuperQPlus100, and from the Swedish Foundation for Strategic Research.

## A Derivation of the average gate fidelity

In this supplementary material, we expand on Ref. [41] to obtain the fidelity correction for $N$-qubit gates going outside of the computational subspace. We start from Eq. (5) in the main text to obtain the expression for $\delta F_N(\hat{L}(t))$ given in Eq. (6) in the main text.

A general $N$-qubit density matrix can be written as

$$\hat{\rho} = \frac{1}{d}\left(\hat{1}_N + \sum_{i=1}^{d^2-1} c_i \hat{f}_i\right) \equiv \frac{1}{d}\left(\hat{\sigma}_0^1 \dots \hat{\sigma}_0^N + \sum_{i_1,\dots,i_N} c_{i_1\dots i_N} \hat{\sigma}_{i_1}^1 \dots \hat{\sigma}_{i_N}^N\right), \tag{20}$$

where the $\hat{f}_i$ consist of tensor products of Pauli matrices, $d = 2^N$, and the $N$ indices $i_1 \dots i_N$ are collected into the single combined index $1 \le i \le d^2 - 1$. We note that the $\hat{f}_i$ basis excludes the identity element corresponding to the term $i_1 = i_2 = \dots = i_N = 0$. Following the symmetry arguments given in Ref. [62], we have [41]

$$\langle c_i \rangle = 0, \tag{21}$$

$$\langle c_i c_j \rangle = \delta_{ij}/(d+1), \tag{22}$$

where the expectation value is taken over all the initial states, i.e., performing the integral $\int d\psi$.

Inserting Eq. (20) into Eq. (5) we obtain

$$\delta F_N(\hat{L}(t)) = \frac{1}{d^2}\int d\psi\, \mathrm{Tr}_{\mathrm{cmp}}\left[\hat{L}^\dagger(t)\left(\hat{\sigma}_0^1\dots\hat{\sigma}_0^N + \sum_{i_1,\dots,i_N} c_{i_1\dots i_N}\hat{\sigma}_{i_1}^1\dots\hat{\sigma}_{i_N}^N\right)\right.$$
$$\left.\times\hat{L}(t)\left(\hat{\sigma}_0^1\dots\hat{\sigma}_0^N + \sum_{j_1,\dots,j_N} c_{j_1\dots j_N}\hat{\sigma}_{j_1}^1\dots\hat{\sigma}_{j_N}^N\right)\right]$$
$$-\frac{1}{d}\int d\psi\, \mathrm{Tr}_{\mathrm{cmp}}\left[\hat{L}^\dagger(t)\hat{L}(t)\left(\hat{\sigma}_0^1\dots\hat{\sigma}_0^N + \sum_{i_1,\dots,i_N} c_{i_1\dots i_N}\hat{\sigma}_{i_1}^1\dots\hat{\sigma}_{i_N}^N\right)\right]. \tag{23}$$

where "cmp" marks that the trace is taken over the states in the computational subspace. Note that as the unitary operation takes the state outside of the computational subspace, we have to project both $\hat{L}(t)$, $\hat{L}^\dagger(t)$, and $\hat{L}^\dagger(t)\hat{L}(t)$ onto the computational subspace. Using the relations in Eqs. (21)–(22), averaging over all possible initial states $|\psi\rangle$ reduces Eq. (23) to

$$\delta F_N(\hat{L}(t)) = \frac{1}{d^2}\mathrm{Tr}_{\mathrm{cmp}}\left[\hat{L}^\dagger(t)\hat{\mathbb{1}}_{\mathrm{cmp}}\hat{L}(t)\hat{\mathbb{1}}_{\mathrm{cmp}}\right] - \frac{1}{d}\mathrm{Tr}_{\mathrm{cmp}}\left[\hat{L}^\dagger(t)\hat{L}(t)\right]$$
$$+\frac{1}{d^2(d+1)}\sum_{i_1,\dots,i_N\neq 0}\mathrm{Tr}_{\mathrm{cmp}}\left[\hat{L}^\dagger(t)\left(\hat{\sigma}_{i_1}^1\dots\hat{\sigma}_{i_N}^N\right)\hat{L}(t)\left(\hat{\sigma}_{i_1}^1\dots\hat{\sigma}_{i_N}^N\right)\right], \tag{24}$$

where we define the identity operation acting only in the computational subspace, $\hat{\mathbb{1}}_{\mathrm{cmp}} = \hat{\sigma}_0^1\dots\hat{\sigma}_0^N \oplus \hat{0}$, where $\hat{0}$ is a zero matrix in the space outside the computational subspace. We note that since $\hat{L}(t)$ might not be confined to the computational subspace, we have, in general, $\mathrm{Tr}_{\mathrm{cmp}}\left[\hat{L}^\dagger(t)\hat{\mathbb{1}}_{\mathrm{cmp}}\hat{L}(t)\hat{\mathbb{1}}_{\mathrm{cmp}}\right] \neq \mathrm{Tr}_{\mathrm{cmp}}\left[\hat{L}^\dagger(t)\hat{L}(t)\right]$.

Equation (24) can be rewritten as

$$\delta F_N(\hat{L}(t)) = \frac{1}{d(d+1)}\mathrm{Tr}_{\mathrm{cmp}}\left[\hat{L}^\dagger(t)\hat{\mathbb{1}}_{\mathrm{cmp}}\hat{L}(t)\hat{\mathbb{1}}_{\mathrm{cmp}}\right] - \frac{1}{d}\mathrm{Tr}_{\mathrm{cmp}}\left[\hat{L}^\dagger(t)\hat{L}(t)\right]$$
$$+\frac{1}{d^2(d+1)}\sum_{i_1,\dots,i_N}\mathrm{Tr}_{\mathrm{cmp}}\left[\hat{L}^\dagger(t)\left(\hat{\sigma}_{i_1}^1\dots\hat{\sigma}_{i_N}^N\right)\hat{L}(t)\left(\hat{\sigma}_{i_1}^1\dots\hat{\sigma}_{i_N}^N\right)\right], \tag{25}$$

which can be written as Eq. (6) in the main text.

## B Fidelity correction for N-qubit gates

Here we present the details of the derivation of the result in Eq. (7) in the main text, which quantifies the fidelity reduction in an $N$-qubit system. We start by calculating $\sum_i \mathrm{Tr}\left[\hat{f}_j\,\hat{f}_i\,\hat{f}_k\,\hat{f}_i\right]$. To do so, we write $\hat{f}_i$ in the basis of Pauli matrices as $\hat{f}_i \equiv f_{i_1+4\,i_2+16\,i_3+\dots+4^{N-1}\,i_N} = \hat{\sigma}_{i_1}^1\hat{\sigma}_{i_2}^2\cdots\hat{\sigma}_{i_N}^N$, with $i_1, i_2, \cdots, i_N = 0, 1, 2, 3$, and obtain

$$\sum_{i=0}^{d^2-1}\mathrm{Tr}\left[\hat{f}_j\,\hat{f}_i\,\hat{f}_k\,\hat{f}_i\right] = \sum_{i_1,\cdots,i_N}^{3}\mathrm{Tr}\left[\left(\hat{\sigma}_{j_1}^1\cdots\hat{\sigma}_{j_N}^N\right)+\left(\hat{\sigma}_{i_1}^1\cdots\hat{\sigma}_{i_N}^N\right)\left(\hat{\sigma}_{k_1}^1\cdots\hat{\sigma}_{k_2}^N\right)\left(\hat{\sigma}_{i_1}^1\cdots\hat{\sigma}_{i_N}^N\right)\right]$$
$$=\sum_{i_1=0}^{3}\mathrm{Tr}\left[\hat{\sigma}_{j_1}\hat{\sigma}_{i_1}\,\hat{\sigma}_{k_1}\hat{\sigma}_{i_1}\right]\cdots\times\sum_{i_N=0}^{3}\mathrm{Tr}\left[\hat{\sigma}_{j_N}\hat{\sigma}_{i_N}\,\hat{\sigma}_{k_N}\hat{\sigma}_{i_N}\right]. \tag{26}$$

We proceed to calculation one summation, $\sum_{i_1=0}^{3}\mathrm{Tr}\left[\hat{\sigma}_{j_1}\hat{\sigma}_{i_1}\,\hat{\sigma}_{k_1}\hat{\sigma}_{i_1}\right]$; the others are the same. We first note that terms with $j_1 \neq k_1$ are traceless, because having two Pauli matrices with the same index leaves a single Pauli matrix that is traceless. For example, considering $j_1 = 0$, we obtain $4\,\mathrm{Tr}\left[\hat{\sigma}_{k_1}\right] = 0$ as $k_1 \neq 0$. If $j_1 \neq k_1 \neq 0$, as $i_1$ can be either $j_1$ or $k_1$, we are left with $\mathrm{Tr}\left[\hat{\sigma}_{k_1}\right] = 0$ or $\mathrm{Tr}\left[\hat{\sigma}_{j_1}\right] = 0$, respectively. This lets us calculate

$$\sum_{i_1=0}^{3} \mathrm{Tr}\left[\hat{\sigma}_{j_1}\hat{\sigma}_{i_1}\,\hat{\sigma}_{j_1}\hat{\sigma}_{i_1}\right] = 8\,\delta_{j_1 0} + \sum_{i_1=0}^{3} \mathrm{Tr}\left[\hat{\sigma}_{j_1\neq 0}\,\hat{\sigma}_{i_1}\,\hat{\sigma}_{j_1\neq 0}\,\hat{\sigma}_{i_1}\right] = 8\,\delta_{j_1 0} + 2 + \sum_{i_1=1}^{3} \mathrm{Tr}\left[\hat{\sigma}_{j_1\neq 0}\,\hat{\sigma}_{i_1}\,\hat{\sigma}_{j_1\neq 0}\,\hat{\sigma}_{i_1}\right]. \tag{27}$$

We use $\hat{\sigma}_i\hat{\sigma}_j = \delta_{ij}\hat{1}_2 + i\epsilon_{ijk}\hat{\sigma}_k$ to calculate the last term and find

$$\sum_{i_1=0}^{3} \mathrm{Tr}\left[\hat{\sigma}_{j_1}\hat{\sigma}_{i_1}\,\hat{\sigma}_{j_1}\hat{\sigma}_{i_1}\right] = 8\,\delta_{j_1 0}+2+\sum_{i_1=1}^{3}\left(\delta_{i_1 j_1\neq 0} - \epsilon^2_{i_1\, j_1\neq 0\, k_1}\right)\mathrm{Tr}\left[\hat{1}_2\right] = 8\,\delta_{j_1 0}+2+(-1)\times 2 = 8\,\delta_{j_1 0}, \tag{28}$$

where we use that $\epsilon^2_{ijk} = 1$ if all indices are different, and 0 otherwise. Equation (26) reduces to

$$\sum_{i=0}^{d^2-1} \mathrm{Tr}\left[\hat{f}_j\,\hat{f}_i\,\hat{f}_k\,\hat{f}_i\right] = 8^N \times \delta_{j_1=k_1 0}\cdots\delta_{j_N=k_N 0} \equiv d^3\delta_{j0}\delta_{k0}. \tag{29}$$

This means the only non-zero terms of the summation in Eq. (6) is the contribution of the identity $\hat{f}_0 \equiv \hat{1}_N$ in $\hat{L}(t)$.

Here we note that $\hat{f}_i$ has no support outside the computational subspace, so terms confined to the computational subspace in $\hat{L}(t)$ have only contributed to $\mathrm{Tr}_{\mathrm{cmp}}\left[\hat{L}^\dagger(t)\,\hat{f}_i\,\hat{L}(t)\,\hat{f}_i\right]$. Expanding those terms in the $\hat{f}_i$ basis, and using Eq. (29), we write the jump operator as

$$\hat{L}(t) = \frac{1}{d}\,\mathrm{Tr}_{\mathrm{cmp}}\left[\hat{L}(t)\right]\hat{f}_0 + \mathcal{O}(\hat{f}_1,\cdots,\hat{f}_{d^2-1}), \tag{30}$$

and obtain

$$\begin{aligned}\sum_{i=0}^{d^2-1} \mathrm{Tr}_{\mathrm{cmp}}\left[\hat{L}^\dagger(t)\,\hat{f}_i\,\hat{L}(t)\,\hat{f}_i\right] &= \frac{1}{d^2}\,\mathrm{Tr}_{\mathrm{cmp}}\left[\hat{L}^\dagger(t)\right]\mathrm{Tr}_{\mathrm{cmp}}\left[\hat{L}(t)\right]\sum_{i=0}^{d^2-1}\mathrm{Tr}_{\mathrm{cmp}}\left[\hat{f}_0\,\hat{f}_i\,\hat{f}_0\,\hat{f}_i\right] \\ &= d\,\mathrm{Tr}_{\mathrm{cmp}}\left[\hat{L}^\dagger(t)\right]\mathrm{Tr}_{\mathrm{cmp}}\left[\hat{L}(t)\right].\end{aligned} \tag{31}$$

Together with Eq. (6), this yields the fidelity reduction for the $N$-qubit system:

$$\begin{aligned}\delta F_N(\hat{L}(t)) = &\frac{1}{d(d+1)}\mathrm{Tr}_{\mathrm{cmp}}\left[\hat{L}^\dagger(t)\hat{\mathbb{1}}_{\mathrm{cmp}}\hat{L}(t)\hat{\mathbb{1}}_{\mathrm{cmp}}\right] - \frac{1}{d}\mathrm{Tr}_{\mathrm{cmp}}\left[\hat{L}^\dagger(t)\hat{L}(t)\right] \\ &+ \frac{1}{d(d+1)}\,\mathrm{Tr}_{\mathrm{cmp}}\left[\hat{L}^\dagger(t)\right]\mathrm{Tr}_{\mathrm{cmp}}\left[\hat{L}(t)\right],\end{aligned} \tag{32}$$

which is Eq. (7) in the main text.

## C Average gate fidelity for the CZ gate with superconducting qubits

Here we present the details of the derivation of the result in Eq. (12) in the main text. The CZ gate is activated by coupling between $|11\rangle$ and $|20\rangle$, given by the Hamiltonian

$$\hat{H}_{\mathrm{CZ}} = \lambda\,(|11\rangle\langle 20| + |20\rangle\langle 11|). \tag{33}$$

The unitary operation $U_{\mathrm{CZ}}(t) = \exp[-iH_{\mathrm{CZ}}t]$ becomes

$$\hat{U}_{\mathrm{CZ}}(t) = |00\rangle\langle 00| + |01\rangle\langle 01| + |10\rangle\langle 10| + \cos(\lambda t)\,(|11\rangle\langle 11| + |20\rangle\langle 20|) - i\sin(\lambda t)\,(|11\rangle\langle 20| + |20\rangle\langle 11|), \tag{34}$$

which at time $\tau = \pi/\lambda$ adds a phase factor of $-1$ to $|11\rangle$.

We start with the effect of energy relaxation on qubit 1 on the average gate fidelity. The jump operator is given by

$$\hat{L}_{-}^{q1} = \left(\hat{\sigma}_{01}^{-} + \sqrt{2}\hat{\sigma}_{12}^{-}\right) \otimes \hat{1}, \tag{35}$$

which together with Eq. (34) and

$$\hat{L}_{-}^{q1}(t) = U_{\mathrm{CZ}}^{\dagger}(t)\hat{L}_{-}^{q1}U_{\mathrm{CZ}}(t), \tag{36}$$

leads to

$$\hat{L}_{-}^{q1}(t) = |00\rangle\langle 10| + \cos(\lambda t)\Big(|01\rangle\langle 11| + \sqrt{2}|10\rangle\langle 20|\Big) - i\sin(\lambda t)\Big(|01\rangle\langle 20| + \sqrt{2}|10\rangle\langle 11|\Big). \tag{37}$$

We then easily obtain

$$\mathrm{Tr}_{\mathrm{cmp}}\left[\hat{L}_{-}^{q1\dagger}(t)\,\hat{L}_{-}^{q1}(t)\right] = \mathrm{Tr}\left[|10\rangle\langle 10| + \left(1+\sin^2(\lambda t)\right)|11\rangle\langle 11|\right] = 2 + \sin^2(\lambda t) \tag{38}$$

and

$$\mathrm{Tr}_{\mathrm{cmp}}\left[\hat{L}_{-}^{q1\dagger}(t)\,\hat{\mathbb{1}}_{\mathrm{cmp}}\,\hat{L}_{-}^{q1}(t)\hat{\mathbb{1}}_{\mathrm{cmp}}\right] = \mathrm{Tr}\left[|10\rangle\langle 10| + \left(1+\sin^2(\lambda t)\right)|11\rangle\langle 11|\right] = 2 + \sin^2(\lambda t). \tag{39}$$

Plugging this into Eq. (7) leads to

$$\delta F_{\mathrm{CZ}}(\hat{L}_{-}^{q1}(t)) = -\frac{1}{5}\Big[2 + \sin^2(\lambda t)\Big]. \tag{40}$$

Finally, performing the time integral we find

$$\int_0^{\tau} dt\,\delta F_{\mathrm{CZ}}(\hat{L}_{-}^{q1}(t)) = -\frac{1}{2}\tau + \frac{\sin(2\lambda\tau)}{20\lambda}. \tag{41}$$

Plugging $\tau = \pi/\lambda$ into this result leads to the expression below Eq. (11) in the main text.

The time-dependent jump operation, when the second qubit is affected by relaxation, is given by

$$\hat{L}_{-}^{q2}(t) = U_{\mathrm{CZ}}^{\dagger}(t)\hat{L}_{-}^{q2}U_{\mathrm{CZ}}(t) = |00\rangle\langle 01| + \cos(\lambda t)|10\rangle\langle 11| + i\sin(\lambda t)|10\rangle\langle 20|. \tag{42}$$

Plugging

$$\mathrm{Tr}_{\mathrm{cmp}}\left[\hat{L}_{-}^{q2\dagger}(t)\,\hat{L}_{-}^{q2}(t)\right] = \mathrm{Tr}\left[|01\rangle\langle 01| + \cos^2(\lambda t)|11\rangle\langle 11|\right] = 1 + \cos^2(\lambda t) \tag{43}$$

and

$$\mathrm{Tr}_{\mathrm{cmp}}\left[\hat{L}_{-}^{q2\dagger}(t)\,\hat{\mathbb{1}}_{\mathrm{cmp}}\,\hat{L}_{-}^{q2}(t)\hat{\mathbb{1}}_{\mathrm{cmp}}\right] = \mathrm{Tr}\left[|01\rangle\langle 01| + \cos^2(\lambda t)|11\rangle\langle 11|\right] = 1 + \cos^2(\lambda t) \tag{44}$$

into Eq. (7) leads to

$$\delta F_{\mathrm{CZ}}(\hat{L}_{-}^{q2}(t)) = -\frac{1}{5}\left(1 + \cos^2(\lambda t)\right), \tag{45}$$

and finally, we obtain

$$\int_0^{\tau} dt\,\delta F_{\mathrm{CZ}}(\hat{L}_{-}^{q2}(t)) = -\frac{3}{10}\tau - \frac{\sin(2\lambda\tau)}{20\lambda}. \tag{46}$$

The jump operator for dephasing on qubit 1 is given by

$$\hat{L}_{\phi}^{q1}(t) = |10\rangle\langle 10| + \Big[1+\sin^2(\lambda t)\Big]|11\rangle\langle 11| + \Big[1+\cos^2(\lambda t)\Big]|20\rangle\langle 20| - i\sin(\lambda t)\cos(\lambda t)(|20\rangle\langle 11| - |11\rangle\langle 20|). \tag{47}$$

Note that Eq. (47) is not traceless; projecting it onto the computational subspace by neglecting terms including $|2\rangle$, we find

$$\mathrm{Tr}_{\mathrm{cmp}}\left[\hat{L}_{\phi}^{q1\dagger}(t)\right] = \mathrm{Tr}_{\mathrm{cmp}}\left[\hat{L}_{\phi}^{q1}(t)\right] = 2 + \sin^2(\lambda t), \tag{48}$$

$$\begin{aligned}\mathrm{Tr}_{\mathrm{cmp}}\left[\hat{L}_{\phi}^{q1\dagger}(t)\,\hat{L}_{\phi}^{q1}(t)\right] &= \mathrm{Tr}\left[|10\rangle\langle 10| + \left(\left(1+\sin^2(\lambda t)\right)^2 + \sin^2(\lambda t)\cos^2(\lambda t)\right)|11\rangle\langle 11|\right] \\ &= \frac{1}{2}[7 - 3\cos(2\lambda t)],\end{aligned} \tag{49}$$

$$\mathrm{Tr}_{\mathrm{cmp}}\left[\hat{L}_{\phi}^{q1\dagger}(t)\,\hat{\mathbb{1}}_{\mathrm{cmp}}\,\hat{L}_{\phi}^{q1}(t)\hat{\mathbb{1}}_{\mathrm{cmp}}\right] = \mathrm{Tr}\left[|10\rangle\langle 10| + \left(1+\sin^2(\lambda t)\right)^2|11\rangle\langle 11|\right] = 1 + \Big[1+\sin^2(\lambda t)\Big]^2. \tag{50}$$

Therefore Eq. (7) yields

$$\delta F_{\rm CZ}(\hat{L}_\phi^{q1}(t)) = \frac{1}{20}\Big[2+\sin^2(\lambda t)\Big]^2 + \frac{1}{20}\Big[1+\Big[1+\sin^2(\lambda t)\Big]^2\Big] - \frac{1}{4}\times\frac{1}{2}[7-3\cos(2\lambda t)], \tag{51}$$

leading to

$$\int_0^\tau dt\, \delta F_{\rm CZ}(\hat{L}_\phi^{q1}(t)) = -\frac{31}{80}\tau + \frac{7\sin(2\lambda\tau)}{80\,\lambda} + \frac{\sin(4\lambda\tau)}{320\,\lambda}. \tag{52}$$

When the dephasing process acts on qubit 2, the jump operator is given by

$$\hat{L}_\phi^{q2}(t) = |01\rangle\langle 01| + \cos^2(\lambda t)|11\rangle\langle 11| + i\sin(\lambda t)\cos(\lambda t)(|20\rangle\langle 11| - |11\rangle\langle 20|) + \sin^2(\lambda t)|20\rangle\langle 20|, \tag{53}$$

leading to

$$\begin{aligned}\mathrm{Tr}\left[\hat{L}_\phi^{q2\dagger}(t)\,\hat{L}_\phi^{q2}(t)\right] &= \mathrm{Tr}\left[|01\rangle\langle 01| + \Big(\cos^4(\lambda t)+\sin^2(\lambda t)\cos^2(\lambda t)\Big)|11\rangle\langle 11|\right] \\ &= \frac{1}{2}[3+\cos(2\lambda t)], \end{aligned}\tag{54}$$

$$\mathrm{Tr}\left[\hat{L}_\phi^{q2\dagger}(t)\right] = \mathrm{Tr}\left[\hat{L}_\phi^{q2}(t)\right] = 1+\cos^2(\lambda t), \tag{55}$$

$$\mathrm{Tr}_{\rm cmp}\left[\hat{L}_\phi^{q2\dagger}(t)\,\hat{\mathbb{1}}_{\rm cmp}\,\hat{L}_\phi^{q2}(t)\,\hat{\mathbb{1}}_{\rm cmp}\right] = \mathrm{Tr}\left[|01\rangle\langle 01| + \cos^4(\lambda t)|11\rangle\langle 11|\right] = 1+\cos^4(\lambda t), \tag{56}$$

such that, together with Eq. (7), we find

$$\delta F_{\rm CZ}(\hat{L}_\phi^{q2}(t)) = \frac{1}{20}\Big[1+\cos^2(\lambda t)\Big]^2 + \frac{1}{20}\Big[1+\cos^4(\lambda t)\Big] - \frac{1}{4}\times\frac{1}{2}[3+\cos(2\lambda t)]. \tag{57}$$

Integrating over time leads to

$$\int_0^\tau dt\, \delta F_{\rm CZ}(\hat{L}_\phi^{q2}(t)) = -\frac{3}{16}\tau - \frac{\sin(2\lambda\tau)}{80\,\lambda} + \frac{\sin(4\lambda\tau)}{320\,\lambda}. \tag{58}$$

Adding up the contributions of the decoherence processes treated above, i.e., Eq. (41), Eq. (46), Eq. (52), and Eq. (58), we find that the fidelity for an imperfect two-qubit CZ gate, with arbitrary strength $\lambda$, is given by

$$\begin{aligned}\overline{F}_{\rm CZ} = 1 &- \left[\frac{1}{2}\tau - \frac{\sin(2\lambda\tau)}{20\lambda}\right]\Gamma_1^{q1}\,\tau - \left[\frac{3}{10}\tau + \frac{\sin(2\lambda\tau)}{20\lambda}\right]\Gamma_1^{q2}\,\tau \\ &- \left[\frac{31}{80}\tau - \frac{7\sin(2\lambda\tau)}{80\,\lambda} - \frac{\sin(4\lambda\tau)}{320\,\lambda}\right]\Gamma_\phi^{q1}\,\tau - \left[\frac{3}{16}\tau + \frac{\sin(2\lambda\tau)}{80\,\lambda} - \frac{\sin(4\lambda\tau)}{320\,\lambda}\right]\Gamma_\phi^{q2}\,\tau. \end{aligned}\tag{59}$$

For the CZ gate without any coherent control error, i.e., $\lambda = \pi/\tau$, the fidelity becomes

$$\overline{F}_{\rm CZ} = 1 - \frac{1}{2}\Gamma_1^{q1}\,\tau - \frac{3}{10}\Gamma_1^{q2}\,\tau - \frac{31}{40}\Gamma_\phi^{q1}\,\tau - \frac{3}{8}\Gamma_\phi^{q2}\,\tau, \tag{60}$$

which is Eq. (12) in the main text.

## D Average gate fidelity for the CZ gate with neutral atoms

Here we give the details for calculating the average gate fidelity of a CZ gate with neutral atoms [Eq. (15) in the main text]. The gate is presented in Ref. [45], applied in Ref. [46], and briefly discussed in the main text. For set of states $\{|01\rangle, |0r\rangle\}$ and $\{|10\rangle, |r0\rangle\}$, with one qubit in $|0\rangle$, the other qubit evolves according to the Hamiltonian in Eq. (13) in the main text, i.e.,

$$\hat{H}_1 = \frac{1}{2}(\Omega|1\rangle\langle r| + \Omega^*|r\rangle\langle 1|) - \Delta|r\rangle\langle r|, \tag{61}$$

and the dynamics of the states $\{|11\rangle, |W\rangle\}$ are given by the Hamiltonian in Eq. (14) in the main text, i.e.,

$$\hat{H}_2 = \frac{\sqrt{2}}{2}(\Omega|11\rangle\langle W| + \Omega^*|W\rangle\langle 11|) - \Delta|W\rangle\langle W|, \tag{62}$$

where

$$|W\rangle = \frac{1}{\sqrt{2}}(|r1\rangle + |1r\rangle), \tag{63}$$

meaning that due to the Rydberg blockade, the transition $|1\rangle \to |r\rangle$ of both atoms never populates $|rr\rangle$, such that we obtain the above superposition and not one of its individual components. The total dimension of the Hilbert space we consider is $3^2 = 9$ [note that the parameter $d$ defined in the main text is $d = 2^N = 4$ as we consider the computational subspace (2 levels for each qubit) when we evaluate the average gate fidelity]. The rest of basis states that describe the system are $\{|00\rangle, |rr\rangle, |D\rangle\}$, where

$$|D\rangle = \frac{1}{\sqrt{2}}(|r1\rangle - |1r\rangle). \tag{64}$$

These states are not affected by the dynamics and remain unchanged. The full Hilbert space is thus spanned by

$$\{|01\rangle, |0r\rangle, |10\rangle, |r0\rangle, |11\rangle, |W\rangle, |00\rangle, |rr\rangle, |D\rangle\}, \tag{65}$$

and the Hamiltonian in this basis order becomes

$$\hat{H} = \hat{H}_1 \oplus \hat{H}_1 \oplus \hat{H}_2 \oplus \hat{0}_3, \tag{66}$$

where $\hat{0}_3$ is a $3 \times 3$ zero matrix. The time evolution describing two global Rydberg pulses of length $\tau$ and detuning $\Delta$ with a laser phase change $\xi$ between pulses is given by

$$\hat{U}(t) = \begin{cases} \hat{U}_1(t), & 0 \leq t < \tau \\ \hat{U}_2(t-\tau)\,\hat{U}_1(\tau), & \tau \leq t \leq 2\tau \end{cases} \tag{67}$$

where

$$\hat{U}_1(t) = e^{-i\hat{H}(\Omega)t}, \tag{68}$$

$$\hat{U}_2(t) = e^{-i\hat{H}(\Omega e^{i\xi})t}. \tag{69}$$

The dominant error mechanism for a Rydberg state is decay to states outside of the computational subspace, which we denote $|\mathbf{O}\rangle$. We therefore consider the jump operator $\hat{L}_r = |\mathbf{O}\rangle\langle r|$ with rate $\Gamma_r$, acting on both atoms individually as $\hat{L}_r^{q1} = \hat{L}_r \otimes \hat{1}$ and $\hat{L}_r^{q2} = \hat{1} \otimes \hat{L}_r$, with the rates $\Gamma_r^{q1}$ and $\Gamma_r^{q2}$, respectively. Since the two atoms are homogeneously coupled from $|1\rangle$ to $|r\rangle$, it is enough to calculate the fidelity reduction due to the Rydberg decay on one qubit; without loss of generality, we perform this calculation for qubit 1. The corresponding jump operator is

$$\hat{L}_r^{q1} = |\mathbf{O}\rangle\langle r| \otimes \hat{1} = |\mathbf{O}0\rangle\langle r0| + |\mathbf{O}1\rangle\langle r1| + |\mathbf{O}r\rangle\langle rr|. \tag{70}$$

The state $|\mathbf{O}\rangle$ is uncoupled to the pulses and remains unchanged by them. We therefore have

$$\mathrm{Tr}_{\mathrm{cmp}}\left[\hat{L}_r^{q1}(t)\right] = \mathrm{Tr}_{\mathrm{cmp}}\left[\hat{L}_r^{q1\dagger}(t)\right] = 0, \tag{71}$$

because the trace over qubit 1 is always zero. It is straightforward to find that

$$\hat{L}_r^{q1\dagger}(t)\hat{L}_r^{q1}(t) = \hat{U}(t)^\dagger\,\hat{L}_r^{q1\dagger}\,\hat{L}_r^{q1}\,\hat{U}(t) = \hat{U}^\dagger(t)[|r0\rangle\langle r0| + |r1\rangle\langle r1| + |rr\rangle\langle rr|]\hat{U}(t), \tag{72}$$

where we use $\hat{U}(t)\hat{U}(t)^\dagger = 1$. We note that $|r1\rangle\langle r1| = \frac{1}{\sqrt{2}}(|W\rangle + |D\rangle)$ in Eq. (72) couples basis states that are in different blocks in the Hamiltonian representation in Eq. (66). The last term in Eq. (72) is $\hat{U}(t)^\dagger|rr\rangle\langle rr|\hat{U}(t) = |rr\rangle\langle rr|$, and for $t < \tau$ we have

$$\hat{U}_1(t)^\dagger|r0\rangle = i\frac{\Omega}{\omega_1}e^{-\frac{i\Delta}{2}t}\sin\left(\frac{\omega_1 t}{2}\right)|10\rangle + e^{-\frac{i\Delta}{2}t}\left[\cos\left(\frac{\omega_1 t}{2}\right) - i\frac{\Delta}{\omega_1}\sin\left(\frac{\omega_1 t}{2}\right)\right]|r0\rangle, \tag{73}$$

$$\hat{U}_1(t)^\dagger|r1\rangle = i\frac{\Omega}{\omega_2}e^{-\frac{i\Delta}{2}t}\sin\left(\frac{\omega_2 t}{2}\right)|11\rangle + \frac{e^{-\frac{i\Delta}{2}t}}{\sqrt{2}}\left[\cos\left(\frac{\omega_2 t}{2}\right) - i\frac{\Delta}{\omega_2}\sin\left(\frac{\omega_2 t}{2}\right)\right]|W\rangle + \frac{1}{\sqrt{2}}|D\rangle, \tag{74}$$

where

$$\omega_1 = \sqrt{\Delta^2 + \Omega^2}, \tag{75}$$

$$\omega_2 = \sqrt{\Delta^2 + 2\Omega^2}. \tag{76}$$

The Rabi frequency of these transitions are different; as we discuss in the main text, we select the first pulse time such that it leads to a perfect transition $|11\rangle \to |W\rangle \to |11\rangle$. According to

$$\hat{U}_1(t)|11\rangle = e^{\frac{i\Delta}{2}t}\left[\cos\left(\frac{\omega_2 t}{2}\right) - i\frac{\Delta}{\omega_2}\sin\left(\frac{\omega_2 t}{2}\right)\right]|11\rangle - i\frac{\sqrt{2}\,\Omega}{\omega_1}e^{\frac{i\Delta}{2}t}\sin\left(\frac{\omega_2 t}{2}\right)|W\rangle, \tag{77}$$

this is guaranteed by the choice $\omega_2\tau = 2\pi$, i.e.,

$$\tau = \frac{2\pi}{\sqrt{\Delta^2 + 2\Omega^2}}, \tag{78}$$

which leads to $\hat{U}_1(\tau)|11\rangle = -e^{\frac{i\Delta}{2}\tau}|11\rangle$. Inserting Eq. (73) and Eq. (74) in Eq. (72), we find for $t < \tau$ that

$$\mathrm{Tr}_{\mathrm{cmp}}\left[\hat{L}_r^{q1\dagger}(t)\hat{L}_r^{q1}(t)\right] = \frac{\Omega^2}{\omega_1^2}\sin\left(\frac{\omega_1 t}{2}\right)^2 + \frac{\Omega^2}{\omega_2^2}\sin\left(\frac{\omega_2 t}{2}\right)^2. \tag{79}$$

The state $|\mathbf{O}\rangle$ is uncoupled to the pulses and remains unchanged by them, so

$$\hat{L}_r^{q1}(t) = \hat{L}_r^{q1}\hat{U}(t). \tag{80}$$

Thus

$$\mathrm{Tr}_{\mathrm{cmp}}\left[\hat{L}_r^{q1\dagger}(t)\,\hat{\mathbb{1}}_{\mathrm{cmp}}\,\hat{L}_r^{q1}(t)\,\hat{\mathbb{1}}_{\mathrm{cmp}}\right] = \mathrm{Tr}_{\mathrm{cmp}}\left[\hat{U}(t)^\dagger\,\hat{L}_r^{q1\dagger}\,\hat{\mathbb{1}}_{\mathrm{cmp}}\,\hat{L}_r^{q1}\hat{U}(t)\,\hat{\mathbb{1}}_{\mathrm{cmp}}\right] = 0. \tag{81}$$

We note that after the second pulse is applied, at $t = 2\tau$, where $\hat{U}(2\tau) = \hat{U}_2(\tau)\hat{U}_1(\tau)$, one can find that

$$\begin{aligned}\hat{U}(2\tau)|11\rangle = &\left(\frac{e^{i\Delta\tau}}{\omega_2^2}\right)\left[\left(1 - e^{i\xi}\right)\Omega^2 + \left(\Delta^2 + \left(1 + e^{i\xi}\right)\Omega^2\right)\cos(\omega_2\tau) - i\omega_2\Delta\sin(\omega_2\tau)\right]|11\rangle \\ &- i\left(\frac{\sqrt{2}\Omega e^{i(\Delta\tau - \frac{\xi}{2})}}{\omega_2^2}\right)\left[\Delta(\cos(\omega_2\tau) - 1)\sin\left(\frac{\xi}{2}\right) + \omega_2\sin(\omega_2\tau)\cos\left(\frac{\xi}{2}\right)\right]|W\rangle.\end{aligned} \tag{82}$$

Using Eq. (78), i.e., $\omega_2\tau = 2\pi$, the state $|11\rangle$ thus receives a total phase of $\phi_{11} = \Delta\tau$:

$$\hat{U}(2\tau)|11\rangle = e^{i\Delta\tau}|11\rangle. \tag{83}$$

Here we note that the dynamical phase accumulated by this process is $\xi$-independent.

The parameter $\xi$ is chosen such that population in $|10\rangle$ returns to that state with an accumulated dynamical phase, $\phi_{10}$. This condition corresponds to [45]

$$e^{-i\xi} = \frac{-\omega_1\cos(\frac{\omega_1 t}{2}) + i\Delta\sin(\frac{\omega_1 t}{2})}{\omega_1\cos(\frac{\omega_1 t}{2}) + i\Delta\sin(\frac{\omega_1 t}{2})}. \tag{84}$$

A CZ gate is obtained by selecting $\phi_{11} = 2\phi_{10} - \pi$, which requires the corresponding numerical values of the relevant parameters $\Delta/\Omega = 0.377371$, $\Omega\tau = 4.29268$, and $\xi = 3.90242$ [45]. With this choice of these parameters, we obtain

$$\hat{U}(2\tau)|10\rangle = e^{3.925\,i}|10\rangle. \tag{85}$$

The same calculation as for Eq. (79) can be done for $\tau \leq t \leq 2\tau$, but we omit explicit expressions here for compactness. Using Eq. (7) and taking the integral over time from 0 to $2\tau$ leads to

$$\int_0^{2\tau} dt\delta F(\hat{L}_r^{q1}(t)) = -\frac{15}{58}. \tag{86}$$

Due to permutation symmetry under exchanging the qubits, from Eq. (3) we obtain

$$\bar{F} = 1 - \frac{15}{58}\left(\Gamma_r^{q1} + \Gamma_r^{q2}\right)\tau, \tag{87}$$

which is Eq. (15) in the main text.

# E Average gate fidelity for the CCZS gate

Here we present the details of the derivation of the average gate fidelity of CCZS gates [Eq. (17) in the main text]. The CCZS$(\theta, \phi, \gamma)$ gate can be written as [37]

$$\mathrm{CCZS}(\theta, \varphi, \gamma) = |0\rangle\langle 0| \otimes \mathbb{I} \otimes \mathbb{I} + |1\rangle\langle 1| \otimes U_{\mathrm{CZS}}(\theta, \varphi, \gamma), \tag{88}$$

where

$$\hat{U}_{\mathrm{CZS}}(\theta, \varphi, \gamma) = \begin{pmatrix} 1 & 0 & 0 & 0 \\ 0 & -e^{i\gamma}\sin^2(\theta/2) + \cos^2(\theta/2) & \frac{1}{2}(1 + e^{i\gamma})e^{-i\varphi}\sin\theta & 0 \\ 0 & \frac{1}{2}(1 + e^{i\gamma})e^{i\varphi}\sin\theta & -e^{i\gamma}\cos^2(\theta/2) + \sin^2(\theta/2) & 0 \\ 0 & 0 & 0 & -e^{i\gamma} \end{pmatrix}, \tag{89}$$

and the parameters $\theta$, $\varphi$, and $\gamma$ are set by the coupling strengths $\lambda_1$, $\lambda_2$ [see Eq. (16) in the main text] and the detuning $\delta$ according to the relations

$$\frac{\lambda_2}{\lambda_1} = \frac{\pi\delta}{\sqrt{4\Omega^2 + \delta^2}}, \tag{90}$$

$$\gamma = -e^{i\varphi}\tan\frac{\theta}{2}, \tag{91}$$

$$\Omega = \sqrt{|\lambda_1|^2 + |\lambda_2|^2}. \tag{92}$$

We evaluate the average gate fidelity for a subclass of CCZS gates: $\lambda_1 = \lambda, \lambda_2 = -\lambda e^{i\varphi}$, $\delta = 0$, and gate time $\tau = \pi/\sqrt{2}\lambda$, for which we obtain $\theta = \pi/2$ and $\gamma = 0$, leading to

$$\hat{U}_{\mathrm{CZS}}(\pi/2, \varphi, 0) = \begin{pmatrix} 1 & 0 & 0 & 0 \\ 0 & 0 & e^{-i\varphi} & 0 \\ 0 & e^{i\varphi} & 0 & 0 \\ 0 & 0 & 0 & -1 \end{pmatrix}. \tag{93}$$

This is a SWAP-like operation on qubits $q_2$ and $q_3$, conditioned on qubit $q_1$ being in its excited state, but adds phase factors to $|101\rangle$, $|110\rangle$, and $|111\rangle$.

Energy relaxation affecting qubit 1 is described by the jump operator

$$\hat{L}_-^{q1} = \left(\hat{\sigma}_{01}^- + \sqrt{2}\hat{\sigma}_{12}^-\right) \otimes \hat{1} \otimes \hat{1}, \tag{94}$$

which together with Eq. (88) leads to

$$\mathrm{Tr}_{\mathrm{cmp}}\left[\hat{L}_{-}^{q1\dagger}(t)\,\hat{L}_{-}^{q1}(t)\right] = 5 - \cos\left(\frac{2\pi t}{\tau}\right), \tag{95}$$

$$\mathrm{Tr}_{\mathrm{cmp}}\left[\hat{L}_{-}^{q1\dagger}(t)\,\hat{\mathbb{1}}_{\mathrm{cmp}}\,\hat{L}_{-}^{q1}(t)\,\hat{\mathbb{1}}_{\mathrm{cmp}}\right] = \frac{1}{4}\left[17 - \cos\left(\frac{4\pi t}{\tau}\right)\right]. \tag{96}$$

As $\mathrm{Tr}\left[\hat{L}_{-}^{q1}(t)\right] = 0$, Eq. (7) in the main text becomes

$$\delta F_{\mathrm{CCZS}}(\hat{L}_{-}^{q1}(t)) = -\frac{1}{8}\left[5 - \cos\left(\frac{2\pi t}{\tau}\right)\right] + \frac{1}{72} \times \frac{1}{4}\left[17 - \cos\left(\frac{4\pi t}{\tau}\right)\right]. \tag{97}$$

Performing the time integral we find

$$\int_0^\tau dt\,\delta F_{\mathrm{CCZS}}(\hat{L}_{-}^{q1}(t)) = -\frac{163}{288}\tau. \tag{98}$$

The jump operator for relaxation in the second qubit is given by

$$\hat{L}_{-}^{q2} = \hat{1} \otimes \hat{\sigma}_{01}^{-} \otimes \hat{1}. \tag{99}$$

This expression does not involve decay from $|2\rangle$ as this state is not involved in the gate. It is straightforward to find

$$\delta F_{\mathrm{CCZS}}(\hat{L}_{-}^{q2}(t)) = -\frac{1}{18}\left[7 + \cos\left(\frac{2\pi t}{\tau}\right)\right], \tag{100}$$

which leads to

$$\int_0^\tau dt\,\delta F_{\mathrm{CCZS}}(\hat{L}_{-}^{q2}(t)) = -\frac{7}{18}\tau. \tag{101}$$

The dephasing on qubits 1 and 2 is described by the jump operators

$$\hat{L}_{\phi}^{q1} = (|1\rangle\langle 1| + 2\,|2\rangle\langle 2|) \otimes \hat{1} \otimes \hat{1}, \tag{102}$$

$$\hat{L}_{\phi}^{q2} = \hat{1} \otimes \hat{\sigma}_z \otimes \hat{1}, \tag{103}$$

with rates $2\Gamma_\phi$ and $\Gamma_\phi/2$, respectively. We find

$$\delta F_{\mathrm{CCZS}}(\hat{L}_{\phi}^{q1}(t)) = -\frac{41}{96} + \frac{7}{36}\cos\left(\frac{2\pi t}{\tau}\right) + \frac{1}{96}\cos\left(\frac{4\pi t}{\tau}\right), \tag{104}$$

$$\delta F_{\mathrm{CCZS}}(\hat{L}_{\phi}^{q2}(t)) = -\frac{85}{96} - \frac{1}{72}\cos\left(\frac{2\pi t}{\tau}\right) + \frac{1}{96}\cos\left(\frac{4\pi t}{\tau}\right) \tag{105}$$

which leads to

$$\int_0^\tau dt\,\delta F_{\mathrm{CCZS}}(\hat{L}_{\phi}^{q1}(t)) = -\frac{41}{96}\tau, \tag{106}$$

$$\int_0^\tau dt\,\delta F_{\mathrm{CCZS}}(\hat{L}_{\phi}^{q2}(t)) = -\frac{85}{96}\tau. \tag{107}$$

Adding up the contributions of the decoherence processes treated above, i.e., Eq. (98), Eq. (101), Eq. (106), and Eq. (107), noting that qubit 3 can be treated in the same way as qubit 2, we find that the total average gate fidelity is

$$\begin{aligned}\overline{F}_{\mathrm{CCZS}} &= 1 - \frac{163}{288}\Gamma_1^{q1}\tau - \frac{7}{18}\left(\Gamma_1^{q2} + \Gamma_1^{q3}\right)\tau - \frac{41}{96}\left(2\Gamma_\phi^{q1}\right)\tau - \frac{85}{96}\left(\frac{1}{2}\left(\Gamma_\phi^{q2} + \Gamma_\phi^{q3}\right)\right)\tau \\ &= 1 - \frac{163}{288}\Gamma_1^{q1}\tau - \frac{7}{18}\left(\Gamma_1^{q2} + \Gamma_1^{q3}\right)\tau - \frac{41}{48}\Gamma_\phi^{q1}\tau - \frac{85}{192}\left(\Gamma_\phi^{q2} + \Gamma_\phi^{q3}\right)\tau,\end{aligned} \tag{108}$$

which is Eq. (17) in the main text.

# F Average gate fidelity for simultaneous CZ gates

Here we present the details of the derivation of the result in Eq. (19) in the main text using the calculation in Appendix C. We consider a four-qubit system where two simultaneous CZ gates are applied, on the pairs $(q_1, q_2)$ and $(q_3, q_4)$. The system has permutation symmetry under exchanging the pairs, so it is enough to compute the fidelity reduction for qubits 1 and 2. We start with the effect of energy relaxation on the average gate fidelity. As we already mentioned, since the jump operators are off-diagonal operations we have

$$\mathrm{Tr}_{\mathrm{cmp}}\left[\hat{L}_-^{q1}(t)\right] = \mathrm{Tr}_{\mathrm{cmp}}\left[\hat{L}_-^{q1\dagger}(t)\right] = 0, \tag{109}$$

$$\mathrm{Tr}_{\mathrm{cmp}}\left[\hat{L}_-^{q2}(t)\right] = \mathrm{Tr}_{\mathrm{cmp}}\left[\hat{L}_-^{q2\dagger}(t)\right] = 0. \tag{110}$$

and from Eq. (38), Eq. (39), Eq. (43), and Eq. (44) we obtain

$$\mathrm{Tr}_{\mathrm{cmp}}\left[\hat{L}_-^{q1\dagger}(t)\,\hat{L}_-^{q1}(t)\right] = \mathrm{Tr}_{\mathrm{cmp}}\left[\hat{L}_-^{q1\dagger}(t)\,\hat{\mathbb{1}}_{\mathrm{cmp}}\,\hat{L}_-^{q1}(t)\,\hat{\mathbb{1}}_{\mathrm{cmp}}\right] = 2 + \sin^2(\lambda t), \tag{111}$$

$$\mathrm{Tr}_{\mathrm{cmp}}\left[\hat{L}_-^{q2\dagger}(t)\,\hat{L}_-^{q2}(t)\right] = \mathrm{Tr}_{\mathrm{cmp}}\left[\hat{L}_-^{q2\dagger}(t)\,\hat{\mathbb{1}}_{\mathrm{cmp}}\,\hat{L}_-^{q2}(t)\,\hat{\mathbb{1}}_{\mathrm{cmp}}\right] = 1 + \cos^2(\lambda t). \tag{112}$$

For dephasing on qubit 1, we have from Eq. (48), Eq. (49), and Eq. (50) that

$$\mathrm{Tr}_{\mathrm{cmp}}\left[\hat{L}_\phi^{q1\dagger}(t)\right] = \mathrm{Tr}_{\mathrm{cmp}}\left[\hat{L}_\phi^{q1}(t)\right] = 2 + \sin^2(\lambda t), \tag{113}$$

$$\mathrm{Tr}_{\mathrm{cmp}}\left[\hat{L}_\phi^{q1\dagger}(t)\,\hat{L}_\phi^{q1}(t)\right] = \frac{1}{2}[7 - 3\cos(2\lambda t)], \tag{114}$$

$$\mathrm{Tr}_{\mathrm{cmp}}\left[\hat{L}_\phi^{q1\dagger}(t)\,\hat{\mathbb{1}}_{\mathrm{cmp}}\,\hat{L}_\phi^{q1}(t)\,\hat{\mathbb{1}}_{\mathrm{cmp}}\right] = 1 + \left[1 + \sin^2(\lambda t)\right]^2, \tag{115}$$

and when the dephasing process acts on qubit 2, we have from Eq. (54), Eq. (55), and Eq. (56) that

$$\mathrm{Tr}\left[\hat{L}_\phi^{q2\dagger}(t)\,\hat{L}_\phi^{q2}(t)\right] = \frac{1}{2}[3 + \cos(2\lambda t)], \tag{116}$$

$$\mathrm{Tr}\left[\hat{L}_\phi^{q2\dagger}(t)\right] = \mathrm{Tr}\left[\hat{L}_\phi^{q2}(t)\right] = 1 + \cos^2(\lambda t), \tag{117}$$

$$\mathrm{Tr}_{\mathrm{cmp}}\left[\hat{L}_\phi^{q2\dagger}(t)\,\hat{\mathbb{1}}_{\mathrm{cmp}}\,\hat{L}_\phi^{q2}(t)\,\hat{\mathbb{1}}_{\mathrm{cmp}}\right] = 1 + \cos^4(\lambda t). \tag{118}$$

Plugging the above results into Eq. (18) in the main text leads to

$$\delta F_2(\hat{L}_-^{q1}(t)) = -\frac{4}{17}\left[2 + \sin^2(\lambda t)\right], \tag{119}$$

$$\delta F_2(\hat{L}_-^{q2}(t)) = -\frac{4}{17}\left[1 + \cos^2(\lambda t)\right], \tag{120}$$

$$\delta F_2(\hat{L}_\phi^{q1}(t)) = \frac{1}{17}\left[2 + \sin^2(\lambda t)\right]^2 + \frac{1}{68}\left[1 + \left[1 + \sin^2(\lambda t)\right]^2\right] - \frac{1}{8}[7 - 3\cos(2\lambda t)], \tag{121}$$

$$\delta F_2(\hat{L}_\phi^{q2}(t)) = \frac{1}{17}\left[1 + \cos^2(\lambda t)\right]^2 + \frac{1}{68}\left[1 + \cos^4(\lambda t)\right] - \frac{1}{8}[3 + \cos(2\lambda t)], \tag{122}$$

and performing the time integrals, at $\lambda = \pi/\tau$, yields

$$\int_0^\tau dt\,\delta F_2(\hat{L}_-^{q1}(t)) = -\frac{10}{17}\tau, \tag{123}$$

$$\int_0^\tau dt\,\delta F_2(\hat{L}_-^{q2}(t)) = -\frac{6}{17}\tau, \tag{124}$$

$$\int_0^\tau dt\,\delta F_2(\hat{L}_\phi^{q1}(t)) = -\frac{245}{544}\tau, \tag{125}$$

$$\int_0^\tau dt\,\delta F_2(\hat{L}_\phi^{q2}(t)) = -\frac{117}{544}\tau. \tag{126}$$

Adding up all the contributions, Eq. (3) in the main text gives that the average gate fidelity for the simultaneous two-qubit CZ gates becomes

$$\overline{F}_{\text{CZ-CZ}} = 1 - \frac{10}{17}\left(\Gamma_1^{q1} + \Gamma_1^{q3}\right)\tau - \frac{6}{17}\left(\Gamma_1^{q2} + \Gamma_1^{q4}\right)\tau - \frac{245}{272}\left(\Gamma_\phi^{q1} + \Gamma_\phi^{q3}\right)\tau - \frac{117}{272}\left(\Gamma_\phi^{q2} + \Gamma_\phi^{q4}\right)\tau, \quad (127)$$

where is Eq. (19) in the main text.